# Uncertainty in the financial market and application to forecast abnormal financial fluctuations

Shige Peng[1,*]   Shuzhen Yang[2,*]   Wenqing Zhang[3]

**Abstract:** The integration and innovation of finance and technology have gradually transformed the financial system into a complex one. Analyses of the causes of abnormal fluctuations in the financial market to extract early warning indicators revealed that most early warning systems are qualitative and causal. However, these models cannot be used to forecast the risk of the financial market benchmark. Therefore, from a quantitative analysis perspective, we focus on the mean and volatility uncertainties of the stock index (benchmark) and then construct three early warning indicators: mean uncertainty, volatility uncertainty, and ALM-G-value at risk. Based on the novel warning indicators, we establish a new abnormal fluctuations warning model, which will provide a short-term warning for the country, society, and individuals to reflect in advance.

## 1. Introduction

Basic innovation and technology are fundamental to maintaining a healthy economy (Brumfiel, 2008). The integration of finance and technology can optimize financial allocation efficiency and satisfy people's diversified demands. Financial innovation has optimized the financial system but also increased financial market instability; thus, the economic system has gradually changed from a classical to a complex system (Arthur, 1999). A complex system has critical thresholds and shifts abruptly from one state to another (Scheffer et al., 2009). Insights into the dynamics of a complex system can be gained by focusing on large fluctuations (Gabaix et al., 2003). The tipping points in financial markets include systemic market crashes and severe depressions (Robert et al., 2008). The accumulation of a series of abnormal financial fluctuations led to the financial crisis, which collapsed the financial system and increased social unemployment; the negative impact was difficult to eliminate in a short time. Therefore, studying abnormal financial fluctuations can help understand complex systems to maintain the stability and sustainability of the financial system (Pettifor, 2020). Scientists insist that establishing an early warning system to detect abnormal financial fluctuations would be necessary when society avoids such fluctuations in the future (Buchanan, 2009).

[1] *Co-corresponding author. Institute of Mathematics, Shandong University, Jinan 250100, China; peng@sdu.edu.cn.

[2] *Co-corresponding author. Shandong University-Zhong Tai Securities Institute for Financial Studies, Shandong University, PR China; yangsz@sdu.edu.cn.

[3] First author. Institute of Mathematics, Shandong University, PR China, (zhangwendy@mail.sdu.edu.cn).

There are three classical early warning systems for abnormal fluctuations in financial markets. The estimations of these systems are mostly linear and simple, although the cause of many fluctuations is nonlinear in reality. Thus, combined with the theory of complex economic systems (Battiston et al., 2016), many models, such as the artificial neural network (Nag and Mitra, 1999), binary recursive tree (SR. Ghosh and AR. Ghosh, 2003), and hybrid causal (Lin et al., 2008) models, have been developed. Most warning systems focus on analyzing the causes of abnormal financial fluctuations to extract early warning indicators and monitor the changes in these indicators. Hinsen (2010) demonstrated that these indicators are overly simplistic in that they ignore the non-quantifiable aspects of the financial market. In addition, these models cannot provide the specific time of an abnormal fluctuation or reflect its impact on the benchmark (stock index).

The stock market is a barometer of the economy and a signal of a country's economic situation. Moreover, stocks represent a company's financial assets and essentially mirror its performance. Hence, the stock market can effectively indicate abnormal fluctuations in financial markets. Therefore, it is important to mine stock market information to warn about abnormal financial fluctuations. We first study the data characteristics of the stock indexes, including the mean of the first-order moment, and the volatility of the second-order moment. Second, we explore stock market risk.

Value at risk system (VaR) is an important risk measurement tool for stock markets. However, the VaR model has some limitations in terms of providing early warning about financial crises. It cannot predict the exact timing of crises (Nocetti, 2006). Additionally, under the assumption of normally distributed returns, the VaR model can only be applied to calm periods (Berger and Missong, 2014). This is because the classical VaR model cannot cover different kinds of uncertainty in the financial market, such as mean and volatility uncertainties. Moreover, in the context of robust statistics, it has been argued that the data satisfy a family of models but not a single model (Huber, 1981; Walley, 1991).

Peng (1997, 2004, 2006, 2008, 2019) originally developed the fundamental rigorous mathematical theory of nonlinear expectation. The nonlinear law of large numbers and nonlinear central limit theorem show that the accumulation of many small mean and volatility uncertainties does not disappear, resulting in model uncertainty. Subsequently, upper and nonlinear normal distributions are developed and used to describe model uncertainty. Therefore, we introduce uncertainty (mean or volatility uncertainties) into the VaR model. Recently, Peng et al. (2023) developed a novel VaR model based on volatility uncertainty: G-VaR model (Peng and Yang, 2022). Based on Jin and Peng's (2016, 2021) $\varphi$-max-mean estimation method, we found an adaptive window that satisfies the convergence result in the sample. Peng et al. (2023) demonstrated that the G-VaR model captures the long-term average loss of risky assets and performs better than the GARCH model. However, the restriction on window choice makes it difficult to apply the G-VaR model to more complex risk characteristics.

To reflect the dynamic, uncertainty, and complexity of the time series, we provide an adaptive learning method for estimating the parameters in the G-VaR model (ALM-G-VaR). Theoretically, the violation rate of the G-VaR model converges to the given risk level with a probability of 1. However, based on historical data, the violation rate of the G-VaR does not stable. Therefore, with a given risk level, our main idea is to adjust the adaptive window according to the observation of the violation rate. If the violation rate is larger than the given risk level, we adjust the degree of volatility uncertainty in the G-VaR parameters by reducing the value of the adaptive window; otherwise, we increase the value of the adaptive window.

We define abnormal financial fluctuations as a two-day downward trend with a stock index smaller than -5%. In general, to construct an early warning system for abnormal financial fluctuations, we develop three indicators: mean uncertainty, volatility uncertainty, and G-VaR. First, we introduce an early warning signal for the mean uncertainty of the stock index. By analyzing the time series characteristics of the stock index, we establish suitable early warning thresholds and issue warning signals by monitoring the absolute and relative values of mean uncertainty, which refers to the lower mean and the difference between the upper and lower means. Subsequently, we examine an early warning indicator for volatility uncertainty and establish an early warning threshold to monitor signals related to upper volatility and the difference between the upper and lower volatilities. Finally, based on volatility uncertainty, we devise a risk assessment benchmark, ALM-G-VaR, and establish three cautionary thresholds. The initial cautionary level is the G-VaR at -0.05, which is intended for the surveillance of abnormal fluctuations in stock prices. The two subsequent indicators signaling potential financial crises are the downward line below or near -0.04 and the last warning line below or near -0.10. We use the 2008 global financial crisis to evaluate the predictive power of the developed early warning system with respect to real abnormal financial fluctuations, and then utilize the classification performance evaluation metrics following further intuitive instructions. To facilitate the comparison, in addition to calculating the precision and miss rate, we also calculate the harmonic mean, $F_1$ of the precision and recall rate. Regarding early warnings about abnormal financial fluctuations, the three indicators possess mutual corroborative capacity, and a holistic assessment can yield preliminary warning signals for impending abnormal fluctuations. The cautionary cues derived from the mean and volatility uncertainties elucidate the potential occurrence of abnormal financial fluctuations in the near future, whereas the specific timing of the abnormal fluctuations can be inferred from the cautionary indications provided by the ALM-G-VaR indicators.

## 2. Results

This study utilizes the nonlinear expectation theory to analyze the stock index (benchmark) in the financial market and explore new early warning systems for detecting abnormal financial fluctuations. First, we construct an early warning index for mean uncertainty and establish appropriate warning thresholds for different financial markets by monitoring the lower mean and the mean differences, which can be predicted in advance. Second, we establish volatility uncertainty, comprising the upper volatility and volatility ratio, to detect abnormal fluctuation events in financial markets. Based on these, we further develop the ALM-G-VaR model, which initially issues an early warning of abnormal stock price fluctuations using the -0.05 threshold; then, we examine the significant downward trend of -0.04 and the ALM-G-VaR value of -0.10 to provide a two-day advance warning for even larger abnormal financial fluctuations and crises. For empirical analysis, five prominent financial market indexes—the Standard & Poor's 500 index and NASDAQ index in the United States, Financial Times Stock Exchange 100 index in the United Kingdom, Frankfurt DAX Index in Germany, and Hushen 300 Index in China—are selected to capture the global financial market risk. For simplification, we refer to these indexes as the S&P500, IXIC, FTSE, GDAXI, and CSI300, respectively. An overview of the paper is presented in Figure 1.

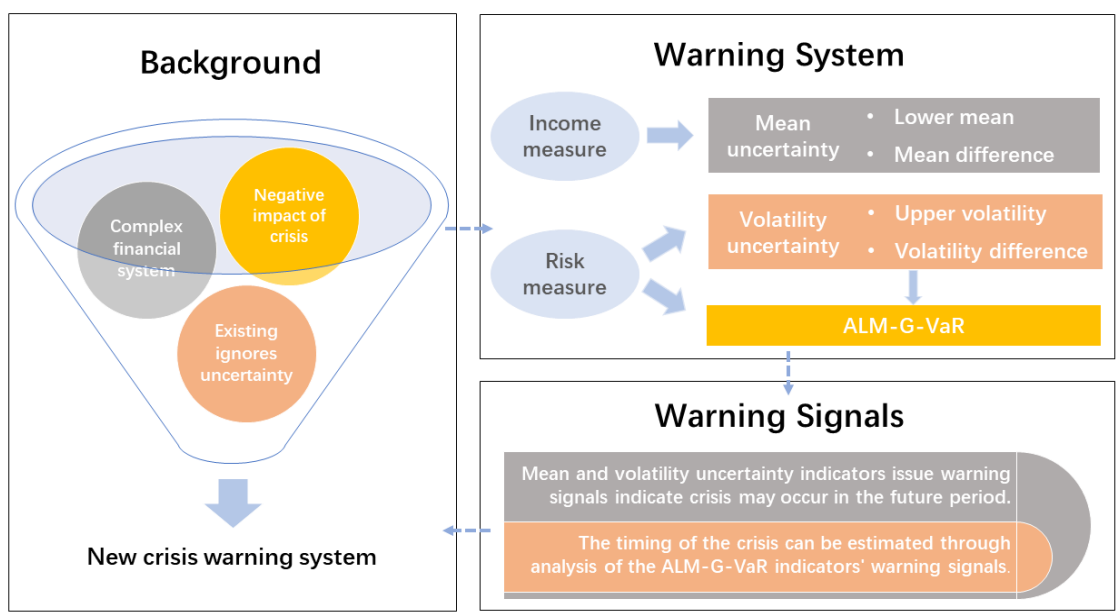

Figure 1 Overview of the paper

To assess the performance of the early warning system established for detecting abnormal financial fluctuations, we first consider the 2008 global financial crisis, a catastrophic event that led to the downfall of global financial markets, as an illustration of its effectiveness. We use stock index returns for two days from January 1, 2007 to January 1, 2010 to predict a financial crisis two days in advance. Subsequently, by expanding the dataset, we further examine the early warning system for abnormal financial fluctuations from January 1, 2008 to June 15, 2023. We can equate the crisis alert to a classification problem, where a positive class can be defined as the occurrence of abnormal financial fluctuations. The specific calculation method is described in Section 4.2. This meticulous examination allows us to make an informed assessment of the precision and effectiveness of these warnings.

**2.1 Time series**

The stock market is the barometer of the economy, a signal of the future economic situation, and is indicative of large financial fluctuations. Therefore, we focus on the stock market, consider the log return of stock data, and extract the changing characteristics from abnormal financial fluctuations. For example, for the S&P500 index from January 1, 2007 to January 1, 2010, there was a cliff-like drop during the 2008 global financial crisis. We set the warning line as -0.05 for the log return, and the warning signal is issued by the log return of S&P500 index for September 23, 2008. As the timing of the financial crisis is based on the stock market crash on September 23, 2008, it can be viewed as the beginning of the 2008 global financial crisis. In 2009, log returns breached the early warning line on several occasions. Based on the above analysis, we explain that a financial meltdown is typically reflected in a significant downturn in stock index time series. By examining the mean uncertainty, volatility uncertainty, and risk measurement ALM-G-VaR, we can extract valuable information to investigate an effective abnormal fluctuation early warning system. To enhance comprehension of the article, we provide pertinent clarifications for certain technical terms mentioned therein, which are given in Table 1. More details on the mean and volatility uncertainties can be found in Section 4.

Table 1 Clarifications for certain technical terms

| Notation | Name | Explanation |
|---|---|---|
| - | Financial abnormal fluctuation | Two consecutive days with the decline in the log return of stock index exceeding 5% |
| $\underline{\mu}$ | Lower mean | Minimum value of mean uncertainty |
| $\bar{\mu}$ | Upper mean | Maximum value of mean uncertainty |
| $\bar{\mu} - \underline{\mu}$ | Mean difference | Difference between the maximum mean and the minimum mean |
| $\underline{\sigma}$ | Lower volatility | Minimum value of volatility uncertainty |
| $\bar{\sigma}$ | Upper volatility | Maximum value of volatility uncertainty |
| $\bar{\sigma}/\underline{\sigma}$ | Volatility ratio | Ratio of the maximum volatility and the minimum volatility |

We employed the precision rate (P), miss rate (M), and harmonic mean ($F_1$) to evaluate the warning performance of mean and volatility uncertainties. The P is the ratio of correct warning samples to the total warning samples.

$$P = \frac{TP}{TP + FP}.$$

M is the proportion of false warning samples to the total number of actual fluctuation samples.

$$M = \frac{FN}{TP + FN}.$$

The recall rate (R) is the proportion of correct warning samples to the total number of actual fluctuation samples.

$$R = \frac{TP}{TP + FN}.$$

$F_1$ can be calculated by harmonic mean of precision and R,

$$\frac{1}{F_1} = \frac{1}{2}\left(\frac{1}{P} + \frac{1}{R}\right).$$

Furthermore, we have

$$F_1 = \frac{2TP}{2TP + FP + FN}.$$

**2.2 Mean uncertainty**

Here, we develop the warning indicators for mean uncertainty, and monitor its values, namely the lower mean and the difference between the upper and lower means (mean difference). Subsequently, we set up appropriate early warning lines to issue crisis signals. The classical mean value fails to reflect the mean aggregation due to the nonlinearity and uncertainty in financial market. Thus, we use the nonlinear expectation theory to describe the nonlinearity and uncertainty of the financial market and assume the log returns of the stock indexes satisfy maximal distribution. Employing the $\varphi$-max-mean parameter estimation method developed by Jin and Peng (2016, 2021), we further use a moving block method to estimate the mean uncertainty. Block lengths $n_0$ and $n_1$ affect the degree of uncertainty; thus, we select the appropriate length by experimenting multiple times with difference sizes of the window. Table 2 presents the specific settings of window length and Section 4.2 presents the method.

Figure 2 shows the classical, upper, and lower mean values of the S&P500 index. A lower mean value denotes the worst-case average return and demonstrates a pronounced downward trend than the classical mean value, implying a smaller return. Hence, we use the lower mean as an early warning indicator and set its lower bound as the warning line. Furthermore, when the difference between the upper and lower mean values is larger, the uncertainty of the mean value is increased. Thus, we set the upper bound as the warning line and monitor it using a crisis warning system. The warning lines vary for different financial markets, and the specific settings are listed in Table 2.

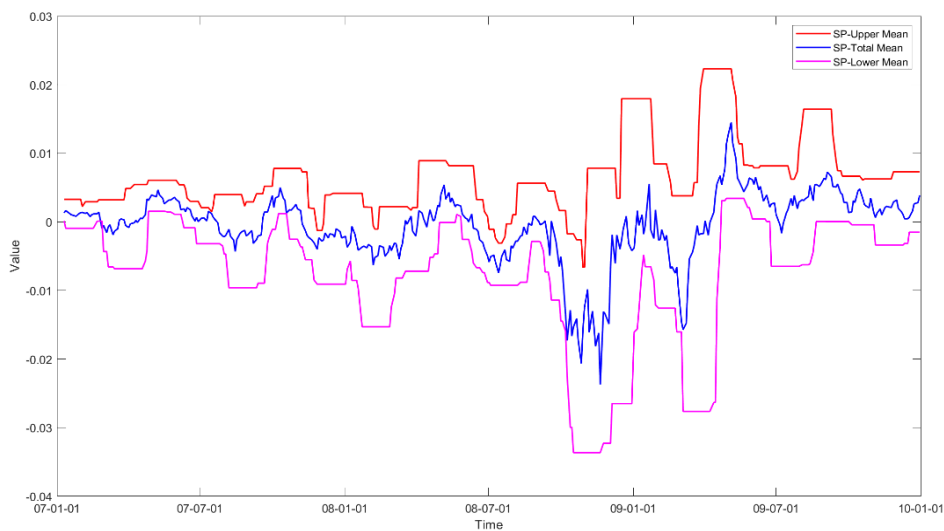

Figure 2 Upper, lower, and classical means of the S&P500 index

Figures 3 and 4 present the mean uncertainty of the S&P500 index. The warning lines are 0.01 for the lower mean and 0.015 for the mean difference. We can observe that the lower mean and mean difference of the S&P500 index triggered the warning lines and issued warning signals on September 19, two days before the crisis. Moreover, on January 23, the mean uncertainty of the S&P500 index sent alerts corresponding to the 2008 global stock disaster. The above observation is also correct for the IXIC and FTSE indexes (see Table 2), which indicates that the mean uncertainty can send out early warning signals two days in advance, and the warning signals of the lower mean and mean difference can mutually support one another. In addition, owing to the delayed influence of Germany and the relatively minimal impact of China during the financial crisis, there was no alarm on the GDAXI and CSI300 indexes mean uncertainties in September 2008. Thus, mean uncertainty can serve as a crisis early warning indicator.

Subsequently, to analyze the precision of mean uncertainty with respect to abnormal stock market fluctuations, we extracted data from January 1, 2008 to June 15, 2023, and closely examined the precision and miss rate. Regarding the accuracy of the lower mean of the S&P500 index, there were 23 warning signals, 14 of which indicated abnormal fluctuations in the real financial market, resulting in a precision of 60.87%. Among the 20 abnormal financial fluctuations, six events were not warned, indicating a miss rate of 30.00%. With respect to the accuracy of the mean difference, the S&P500 index has 68.97% precision and a 0.00% miss rate. By employing the same methodology, we can assess the accuracy of other stock indexes, resulting in an average precision of 63.48% for the lower mean and 67.84% for the mean difference, and an average miss rate of 48.92% for the lower mean and 32.23% for the mean difference. Both values of the $F_1$ mean warning indicators are less than 0.7. Combined with the performance of mean uncertainty during the 2008 financial crisis and the past 15 years of abnormal financial fluctuations, mean uncertainty can serve as an early warning indicator of abnormal financial fluctuations. In contrast, mean uncertainty exhibits delayed responses and underreporting. Therefore, we consider a shift from relying solely on the first-order moment to incorporating the second-order moment–that is, volatility.

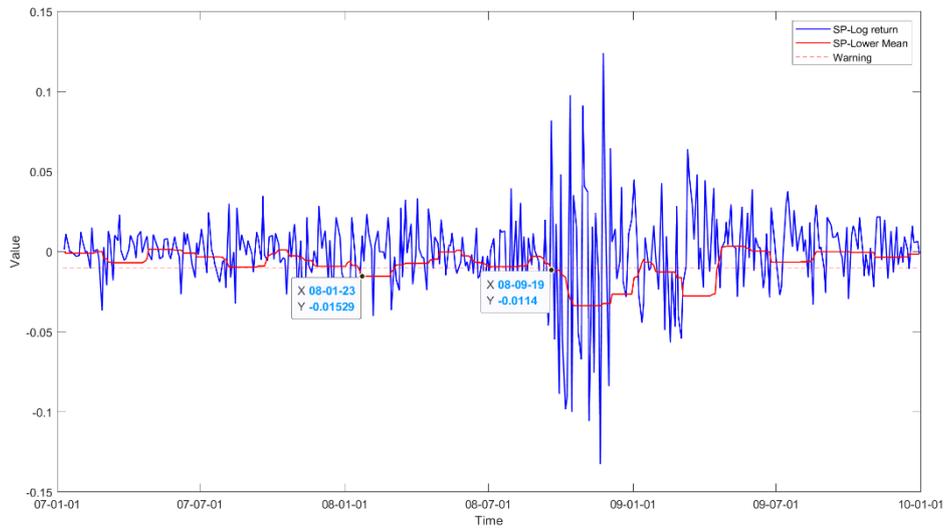

Figure 3 Lower mean and log return of the S&P500 index

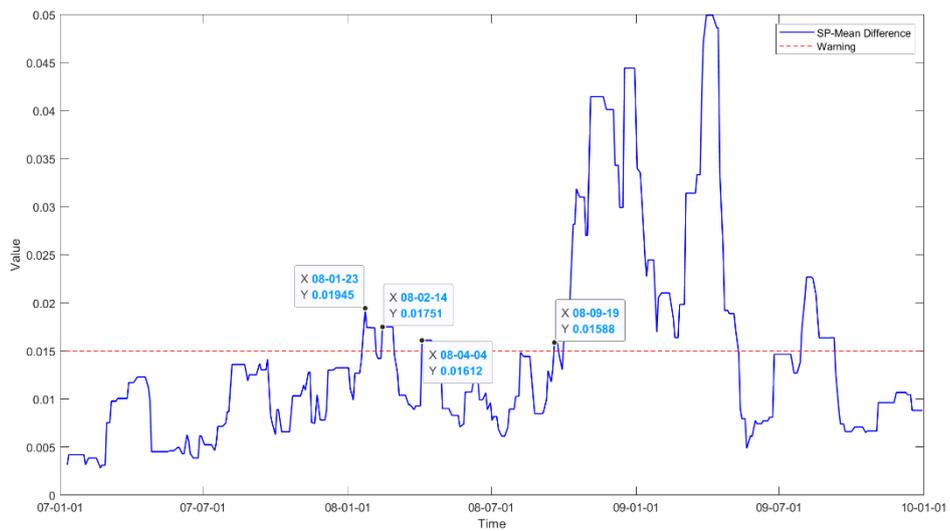

Figure 4 Mean uncertainty of the S&P500 index

Table 2 Lower mean and mean difference of the stock indexes

| Signal | Index | Block | Line | Time | Value | Precision | Miss | $F_1$ |
|---|---|---|---|---|---|---|---|---|
| Lower mean | S&P500 | $n_0$: 20<br>$n_1$: 8 | <-0.01 | 2008.01.23 | -0.0153 | 60.87% | 30.00% | 0.6512 |
| | | | | 2008.09.19 | -0.0114 | | | |
| | IXIC | $n_0$: 20<br>$n_1$: 9 | <-0.013 | 2008.01.23 | -0.0156 | 62.96% | 34.62% | 0.6415 |
| | | | | 2008.09.19 | -0.0140 | | | |
| | FTSE | $n_0$: 20<br>$n_1$: 7 | <-0.015 | 2008.01.23 | -0.0209 | 69.23% | 55.00% | 0.5455 |
| | | | | 2008.03.19 | -0.0167 | | | |
| | | | | 2008.09.19 | -0.0188 | | | |
| | GDAXI | $n_0$: 20 | <-0.02 | 2008.01.23 | -0.0246 | 58.82% | 61.54% | 0.4651 |

| | | | | | | | |
|---|---|---|---|---|---|---|---|
| | | $n_1$: 7 | | 2008.10.13 | -0.0338 | | |
| | CSI300 | $n_0$: 20<br>$n_1$: 5 | <-0.03 | 2008.01.31 | -0.0358 | 65.52% | 63.46% | 0.4691 |
| | | | | 2008.06.16 | -0.0351 | | |
| | | | | 2008.08.13 | -0.0373 | | |
| Mean difference | S&P500 | $n_0$: 20<br>$n_1$: 8 | >0.015 | 2008.01.23 | 0.0195 | 68.97 | 0.00% | 0.8163 |
| | | | | 2008.02.14 | 0.0175 | | |
| | | | | 2008.04.04 | 0.0161 | | |
| | | | | 2008.09.19 | 0.0159 | | |
| | IXIC | $n_0$: 20<br>$n_1$: 9 | >0.02 | 2008.01.23 | 0.0203 | 65.71% | 11.54% | 0.7541 |
| | | | | 2008.09.19 | 0.0209 | | |
| | FTSE | $n_0$: 20<br>$n_1$: 7 | >0.02 | 2008.01.21 | 0.0243 | 70.59% | 40.00% | 0.6486 |
| | | | | 2008.03.19 | 0.0242 | | |
| | | | | 2008.09.19 | 0.0234 | | |
| | GDAXI | $n_0$: 20<br>$n_1$: 7 | >0.03 | 2008.01.23 | 0.0337 | 69.23% | 30.77% | 0.6923 |
| | | | | 2008.04.09 | 0.0315 | | |
| | | | | 2008.10.13 | 0.0348 | | |
| | CSI300 | $n_0$: 20<br>$n_1$: 5 | >0.05 | 2008.01.31 | 0.0590 | 64.71% | 78.85% | 0.3188 |
| | | | | 2008.03.20 | 0.0540 | | |
| | | | | 2008.04.30 | 0.0538 | | |
| | | | | 2008.07.16 | 0.0546 | | |
| | | | | 2008.08.13 | 0.0501 | | |

## 2.3 Volatility uncertainty

Turning to the second-order moment volatility due to individual investment risk preferences, risk assets often exhibit a volatility aggregation phenomenon that cannot be properly described by the classical distribution model. Therefore, we employ the nonlinear expectation theory to demonstrate volatility uncertainty and devise the upper volatility and the ratio between the upper and lower volatility (volatility ratio) for crisis warning. The larger the two indicators, the greater the volatility uncertainty. Thus, we establish their upper bounds as warning lines, which vary for different financial markets. The moving block method was used to estimate volatility uncertainty. We selected suitable block lengths $n_0$ and $n_1$ for different financial markets by experimenting multiple times with different window sizes. The results and calculation methods are presented in Table 3 and Section 4.3, respectively.

As shown in Table 3, the upper volatility and volatility ratio of the S&P500, IXIC,

and FTSE indexes broke the warning lines and sent warning signals on September 23. This indicates that volatility uncertainty can send early warning signals, and the warning signals of the upper volatility and volatility ratio mutually support each other. As Germany and China were less affected by the 2008 financial crisis, the volatility uncertainty of the GDAXI and CSI300 indexes did not trigger warning signals for the 2008 financial crisis. Overall, the volatility uncertainty serves as an early warning indicator. Similar to the accuracy assessment conducted for mean uncertainty, we analyze forecasting effectiveness pertaining to volatility uncertainty. By utilizing a dataset spanning 15 years, we determined that the average precision of volatility uncertainty was 69.54% for the upper volatility and 62.62% for the volatility ratio. In addition, the average miss rate was 26.69% for upper volatility and 72.08% for the volatility ratio. The $F_1$ values of both volatility warning indicators were less than 0.7. This finding suggests a significant presence of false signals within the warnings generated by volatility uncertainty. Furthermore, certain warning signals exhibit a lag, meaning that they are issued only once a crisis has already occurred, which leads to a high miss rate. Therefore, a precise and accurate warning model is required.

Table 3 Upper volatility and volatility ratio of the stock indexes

| Signal | Index | Block | Line | Time | Value | Precision | Miss | $F_1$ |
|---|---|---|---|---|---|---|---|---|
| Upper volatility | S&P500 | $n_0$: 20 $n_1$: 8 | >0.03 | 2008.09.23 | 0.0397 | 69.23% | 10.00% | 0.7826 |
| | IXIC | $n_0$: 20 $n_1$: 9 | >0.03 | 2008.09.23 | 0.0375 | 64.86% | 7.69% | 0.7619 |
| | FTSE | $n_0$: 20 $n_1$: 7 | >0.04 | 2008.09.23 | 0.0476 | 72.22% | 35.00% | 0.6842 |
| | GDAXI | $n_0$: 20 $n_1$: 7 | >0.04 | 2008.01.29 | 0.0447 | 72.41% | 19.23% | 0.7636 |
| | | | | 2008.10.09 | 0.0470 | | | |
| | CSI300 | $n_0$: 20 $n_1$: 5 | >0.05 | 2008.02.13 | 0.0525 | 68.97% | 61.54% | 0.4938 |
| | | | | 2008.07.10 | 0.0519 | | | |
| Volatility ratio | S&P500 | $n_0$: 20 $n_1$: 8 | >3 | 2008.09.23 | 3.0675 | 62.50% | 75.00% | 0.3571 |
| | IXIC | $n_0$: 20 $n_1$: 9 | >2.5 | 2008.09.23 | 2.5612 | 62.50% | 42.31% | 0.6000 |
| | FTSE | $n_0$: 20 $n_1$: 7 | >3 | 2008.02.08 | 3.0033 | 66.67% | 70.00% | 0.4138 |
| | | | | 2008.05.01 | 3.0523 | | | |
| | | | | 2008.07.02 | 3.2299 | | | |

|  |  |  | 2008.09.23 | 3.3076 |  |  |  |
|---|---|---|---|---|---|---|---|
| GDAXI | $n_0$: 20 | >4.5 | 2008.01.29 | 4.5591 | 71.43% | 80.77% | 0.3030 |
|  | $n_1$: 7 |  | 2008.10.09 | 4.9146 |  |  |  |
| CSI300 | $n_0$: 20 | >5 | 2008.01.11 | 6.4155 | 50.00% | 92.31% | 0.1333 |
|  | $n_1$: 5 |  |  |  |  |  |  |

By combining mean and volatility uncertainties, although they breach the warning lines multiple times and send warning signals before the financial crisis, there are some limitations to the crisis warning. The selection of window length is subjective. We manually adjusted the window length through multiple experiments without a reasonable rule system. In addition, the establishment of warning lines lags and is subjective, depending on the specific financial crisis. They are diverse across different financial markets; thus, it is difficult to establish a uniform standard for different financial crises and markets. Furthermore, these indicators cannot provide specific timing for a crisis. Some of the warning signals, like mean uncertainty, are sent two days before the crisis, and some like volatility uncertainty are on the day of the crisis. Therefore, it is difficult to investigate the beginning of the crisis. Finally, the accuracy of the crisis alerts was deficient. Numerous erroneous indicators are present in the warning signals, and many abnormal fluctuations remain unwarned. To overcome these limitations, we propose an accurate and comprehensive predictor of financial crises.

**2.4 ALM-G-VaR**

VaR models are widely used to measure financial market risks. However, classical VaR models cannot measure uncertainty in the financial market, predict the exact crisis timing (Nocetti, 2006), and be applied during storm periods (Berger and Missong, 2014). Recently, Peng et al. (2023) developed a new VaR model based on volatility uncertainty. To overcome the restrictions on window length selection in the G-VaR model, we develop an adaptive learning method to automatically adjust the window length, the ALM-G-VaR model. For further details, see Section 4.4. Based on daily data, we compute the ALM-G-VaR value for a given risk level of 0.05. The results of the S&P500 index are presented in Figure 5, and those of the other indexes are summarized in Table 4.

Figure 5 shows the excellent performance of the ALM-G-VaR model, indicating that the worst-case distribution can capture local and global changes in the log return of the S&P500 index. We establish a threshold of -0.05 as the initial indication of abnormal stock market volatility. By assessing the performance of the G-VaR, we can extract warning signals that signify an impending crisis. From September 19 to September 23, the ALM-G-VaR value of the S&P500 index decreased from -0.0376 to -0.0839, indicating that the return (every two days) of the S&P500 index decreased by approximately 0.0463. The last warning point -0.0839 is close to the level of -0.10. Similarly, the ALM-G-VaR of the IXIC and FTSE indexes also fell off a cliff between September 17 and September 19, with a decrease of 0.0477 and 0.0393, respectively, and the last warning lines were close to -0.10. Hence, we extract two warning signs of the financial crisis as the downward trend is below or near -0.04 and the last warning point is at or near -0.10. Owing to the limited damage suffered by Germany and China's financial systems during the 2008 financial crisis, the ALM-G-VaR of the GDAXI and CSI300 indexes are far from the warning lines in September 2008. Therefore, we can initially issue early warning of abnormal stock price fluctuations using the -0.05 threshold. Subsequently, by examining the significant downward trend of -0.04 and the ALM-G-VaR value of -0.10, we forecast a financial crisis two days in advance.

Furthermore, by using the data for the period of January 1, 2008 to June 15, 2023, we obtained the accuracy of the ALM-G-VaR early warning. Concerning the warning of abnormal stock price fluctuations in the S&P500 index, there were 24 instances in where the -0.05 ALM-G-VaR threshold was surpassed. Among these, 17 cases exhibited abnormal fluctuations and early warnings with a precision rate of 70.83%. Additionally, of the 20 abnormal fluctuations in the stock market, the predictive power of the ALM-G-VaR was limited to 17, implying a false report rate of 15%. Hence, the average precision and miss rate of abnormal stock price movement warnings across the five stock indexes are 66.45% and 12.15%, respectively, and the average value of $F_1$ is 0.7560.

Shifting our focus to financial crisis warnings, the S&P500, IXIC, and FTSE indexes generated warning signals on September 23, 2008 and March 16, 2020, aligned with the events of the financial crisis and the COVID-19 outbreak, respectively. As the German financial market experienced a delayed impact from both events, the GDAXI stock index issued crisis warnings on October 16, 2008 and March 26, 2020. Similarly, the CSI300 index issued a warning on July 15, 2015, corresponding to China's 2015 stock market crisis. Generally, the ALM-G-VaR warning system is more impartial and enhances the precision of early warning systems.

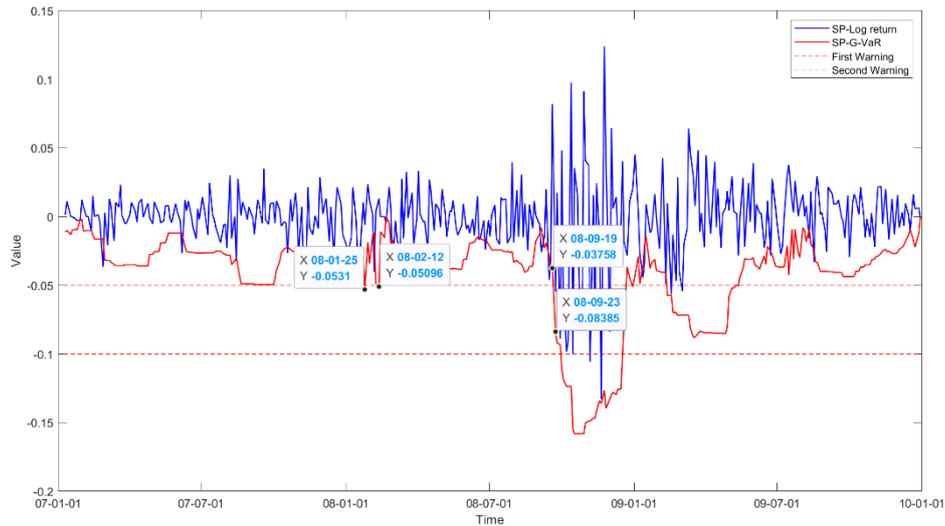

Figure 5 ALM-G-VaR for the S&P500 index from 2007 to 2009

Table 4 ALM-G-VaR warning signal of the stock indexes

| Indexes | Warning signal | Warning time | Value | | Precision rate | Miss rate | $F_1$ |
|---|---|---|---|---|---|---|---|
| S&P500 | ①Value < -0.05 | 2008.01.25 | -0.0531 | | 70.83% | 15.00% | 0.7727 |
| | | 2008.02.12 | -0.0510 | | | | |
| | ②Down Trend < -0.04 | 2008.09.19 | -0.0376 | -0.0463 | 100% | - | - |
| | ③ Second Value < -0.10 | 2008.09.23 | -0.0839 | | | | |
| IXIC | ①Value < -0.05 | 2008.09.23 | -0.0874 | | 64.86% | 7.69% | 0.7619 |
| | ②Down Trend < -0.04 | 2008.09.19 | -0.0397 | -0.0477 | 100% | - | - |
| | ③ Second Value < -0.10 | 2008.09.23 | -0.0874 | | | | |
| FTSE | ①Value < -0.05 | 2008.01.29 | -0.0631 | | 65.38% | 15.00% | 0.7391 |
| | | 2008.09.11 | -0.0547 | | | | |
| | ②Down Trend < -0.04 | 2008.09.19 | -0.0570 | -0.0393 | 100% | - | - |
| | ③ Second Value < -0.10 | 2008.09.23 | -0.0963 | | | | |
| GDAXI | ①Value < -0.05 | 2008.01.23 | -0.0585 | | 69.44% | 3.85% | 0.8065 |
| | ②Down Trend < -0.04 | 2008.09.19 | -0.0478 | -0.0240 | 100% | - | - |

| | | | | | | |
|---|---|---|---|---|---|---|
| | ③ Second Value < -0.10 | 2008.09.23 | -0.0718 | | | |
| CSI300 | ①Value < -0.05 | 2008.01.23 | -0.0513 | | 61.76% | 19.23% | 0.7000 |
| | ②Down Trend < -0.04 | 2008.09.19 | -0.0871 | -0.0049 | 100% | - | - |
| | ③ Second Value < -0.10 | 2008.09.23 | -0.0920 | | | |

By comparing the early warning effects of the three indicators, we can observe that the precision of all three indicators exceeds 60%; however, the miss rate differs among the three indicators. The ALM-G-VaR controls it at approximately 12%, which is significantly lower than that of the mean and variance. Considering the $F_1$ value, the average $F_1$ value of ALM-G-VaR is higher than 0.75, whereas the average $F_1$ values of the mean value and variance are lower than 0.7, which indicates that ALM-G-VaR has a better prediction performance. Moreover, the ALM-G-VaR indicator can distinguish between general and large abnormal fluctuations such as financial crises by establishing different early warning signals. By comparison, it can be concluded that ALM-G-VaR has a lower false reporting rate and a better prediction effect.

A comprehensive analysis of the mean uncertainty, volatility uncertainty, and ALM-G-VaR warning indicators can provide reliable and comprehensive warning signals for financial crises. Based on appropriate early warning lines, a financial crisis may occur when the mean and volatility uncertainty indicators send warning signals. The timing of the crisis can be estimated using the ALM-G-VaR model, which can accurately forecast a financial crisis two days in advance. In conclusion, these warning signals confirm each other and provide valuable insights into financial crises. In addition, the adapted learning method can be used to calculate the mean and volatility uncertainties accurately and timely.

## 3. Discussion

The integration of finance and technology have increased financial allocation efficiency. However, it has also increased financial market uncertainty and risk, which can potentially cause financial crises. Therefore, it is important to establish effective financial crisis warning systems. Most warning systems focus on analyzing the causes of financial crises to extract warning indicators, which oversimplify the complex, non-quantifiable aspects of financial markets (Hinsen, 2010). To address this issue, it is essential to build an early warning system by analyzing the time series of stock indexes. Classical time-series models assume a deterministic data distribution, which is unsuitable for real financial markets that are nonlinear and uncertain. Therefore, a sophisticated approach is required.

Peng (1997, 2004, 2006, 2008, 2019, 2020) originally developed a new rigorous mathematical theory, nonlinear expectation, and integrated it into risk measurement, for example, G-VaR model. By comparing the results of the G-VaR model and conventional risk measurement, the G-VaR model can capture the longtime average loss of risky assets. However, the restriction on the choice of the window makes it difficult to apply the G-VaR model to complex risk characteristics. Hence, we establish an adaptive learning method for estimating the parameter of the G-VaR (ALM-G-VaR) model. In theory, violation rate $\prod(\alpha,t)$ converges to $\alpha$ with a probability of 1, as $t \to \infty$; however, $\prod(\alpha,t)$ is far from $\alpha$ in practice. Therefore, we adjust the estimation of parameters $\left(\hat{\underline{\mu}}_s, \hat{\bar{\mu}}_s, \hat{\underline{\sigma}}_s, \hat{\bar{\sigma}}_s\right)$ by adjusting the data window $(n_0, n_1)$. Based on the sublinear expectation $\mathbb{E}[\cdot]$ and mild conditions, we prove the accuracy of the model; that is, the sublinear expectation of the violation rate converges to $\alpha$ when the amount of financial data tends to infinity, as $\mathbb{E}\left[\tilde{\prod}(\alpha,t)\right] \to \alpha, t \to \infty$.

Therefore, adaptive learning methods can be applied to provide crisis warnings. By considering the return rate of the data and setting a warning line -0.05, it was observed that the stock indexes issued a warning signal on September 23, 2008, which provided early warning information about the arrival of the 2008 global financial crisis. We synthesized three warning signals—mean uncertainty, volatility uncertainty, and ALM-G-VaR—to provide comprehensive and accurate early warning information for the arrival of abnormal financial fluctuations. The results indicate that by setting appropriate warning lines and window lengths, the mean and volatility uncertainties emitted warning signals multiple times before and during the 2008 global financial crisis. Although mean and volatility uncertainties can serve as warning indicators, the lag in issuing warning signals, lack of specificity regarding the timing of crises, and high miss rate pose challenges. To address these issues, we further develop the ALM-G-VaR model for providing early warning of abnormal financial fluctuations, which can effectively forecast an abnormal fluctuation when its value is below -0.05. For an even larger anomaly, when there is a substantial decline in the ALM-G-VaR of stock indexes, such as when the downward trend is below or near -0.04 and the last warning point is at or near -0.10, financial crises can be predicted two days in advance. By combining the three early warning indicators, the warning signals of mean and volatility uncertainties illustrate that abnormal financial fluctuations may occur in the future, and the specific time of the abnormal fluctuations can be determined via the warning signals of the ALM-G-VaR indicators.

The VaR model is widely used for risk measurement and served as the foundation for the ALM-G-VaR warning indicators. However, it has limitations in measuring extreme risk because it only considers a single level of loss probability and lacks a description of tail loss. Additionally, this violates the sub-additivity property proposed by Artzner et al. (1999), indicating that the VaR model is not a consistent risk measure. To address these issues, Acebi and Tasche (2002) developed an accurate risk measurement model, the expected shortfall (ES) model, which describes the average tail loss and is a consistent risk measure. Given the excellent properties of the ES model, the Basel Committee on Banking Supervision released a new, lower capital requirement for market risk, replacing the VaR model with the ES model. As the classical ES model also assumes that financial data follow a deterministic distribution, our future work will incorporate nonlinear expectations with the ES model and utilize the newly developed G-ES model to construct a financial crisis warning system.

## 4. Methods

According to the results given in Section 2, we show the calculation methods in details.

**4.1 Time series**

We perform preprocessing on the stock index data. Let $\{X_s\}$ be the time series of the stock index. The log return of stock index is denoted by $Z_s$,

$$Z_s = \log(X_s) - \log(X_{s-1}) = \log\frac{X_s}{X_{s-1}}. \tag{1}$$

In the following, we consider the log return of the stock index and construct mean uncertainty, volatility uncertainty and ALM-G-VaR warning indicators.

### 4.2 Mean uncertainty

To measure the mean uncertainty of the data, we assume that the mean of $\{Z_s\}_{s=1}^{T}$ satisfy the maximal distribution $M[\underline{\mu}_s, \overline{\mu}_s]$, and are $i.i.d$ under sublinear expectation (see Peng et al. (2019)), where $\underline{\mu}_s$ and $\overline{\mu}_s$ represents the lower and upper means, respectively. Based on the $\varphi$-max-mean parameter estimation method developed in Jin and Peng (2016, 2021), we use a moving block method to estimate the parameters $(\underline{\mu}_s, \overline{\mu}_s)$. With the initially length of historical data $n_0$, considering the data $\{Z_r\}_{s-n_0+1 \leq r \leq s}$, we divide them into $n_0 - n_1 + 1's$ blocks,

$$\{Z_{s-n_0+1}, \cdots, Z_{s-n_0+n_1}\}, \{Z_{s-n_0+2}, \cdots, Z_{s-n_0+n_1+1}\}, \cdots, \{Z_{s-n_1+1}, \cdots, Z_s\},$$

where each block has $n_1$ elements, the sample mean for each moving block is

$$\hat{\mu}_{s,j} = \frac{\sum_{i=1}^{n_1} Z_{s-n_0+i+j-1}}{n_1}, 1 \leq j \leq n_0 - n_1 + 1. \tag{2}$$

The lower and upper means of $\{Z_r\}_{s-n_0+1 \leq r \leq s}$ satisfy

$$\underline{\mu}_s := -\mathbb{E}[-Z_s] \quad and \quad \overline{\mu}_s := \mathbb{E}[Z_s].$$

Applying the $\varphi$-max-mean method in Jin and Peng (2016, 2021), the estimations of lower and upper mean are given by:

$$\hat{\underline{\mu}}_s = \min_{1 \leq j \leq n_0 - n_1 + 1} \hat{\mu}_{s,j} \quad and \quad \hat{\overline{\mu}}_s = \max_{1 \leq j \leq n_0 - n_1 + 1} \hat{\mu}_{s,j}. \tag{3}$$

### 4.3 Volatility uncertainty

In this part, we assume the log return of stock index $Z_s$ without mean uncertainty,

that is $\mathbb{E}[Z_s] = -\mathbb{E}[-Z_s] = \mu_s$. Similar with the subsection 4.2, we set the initially window lengths $n_0$ and $n_1$, where $n_1 \leq n_0$. Dividing the data $\{Z_r\}_{s-n_0+1 \leq r \leq s}$ into $n_0 - n_1 + 1's$ blocks,

$$\{Z_{s-n_0+1}, \cdots, Z_{s-n_0+n_1}\}, \{Z_{s-n_0+2}, \cdots, Z_{s-n_0+n_1+1}\}, \cdots, \{Z_{s-n_1+1}, \cdots, Z_s\}.$$

The sample mean for each moving block is

$$\hat{\mu}_{s,j} = \frac{\sum_{i=1}^{n_1} Z_{s-n_0+i+j-1}}{n_1}, 1 \leq j \leq n_0 - n_1 + 1. \quad (4)$$

The estimation of volatility for each moving block is

$$\hat{\sigma}_{s,j}^2 = \frac{\sum_{i=1}^{n_1} \left(Z_{s-n_0+i+j-1} - \hat{\mu}_{s,j}\right)^2}{n_1 - 1}. \quad (5)$$

Then, the estimations of lower and upper volatility are given by:

$$\hat{\underline{\sigma}}_s^2 = \min_{1 \leq j \leq n_0 - n_1 + 1} \hat{\sigma}_{s,j}^2 \quad and \quad \hat{\bar{\sigma}}_s^2 = \max_{1 \leq j \leq n_0 - n_1 + 1} \hat{\sigma}_{s,j}^2. \quad (6)$$

**4.4 ALM-G-VaR model**

To formulate the model of ALM-G-VaR, we first give a basic assumption for the log return of stock index $\{Z_s\}_{s=1}^{T}$.

**Assumption 4.1.** Let $Z_s$ satisfy a G-normal distribution $N(\mu_s, [\underline{\sigma}_s^2, \bar{\sigma}_s^2])$ under sublinear expectation $\mathbb{E}[\cdot]$, and $Z_s$ be independent from $Z_1, Z_2, \cdots, Z_{s-1}, s \geq 2$, where $(\mu_s, \underline{\sigma}_s, \bar{\sigma}_s)$ are deterministic parameters of G-normal distribution. There exist positive constants $\underline{l}, \bar{l}$ such that

$$\underline{l} < \inf_{s \geq 1} \underline{\sigma}_s, \quad \sup_{s \geq 1} \bar{\sigma}_s < \bar{l},$$

where $\underline{\sigma}_s^2 = -\mathbb{E}[-(Z_s - \mu_s)^2]$ and $\bar{\sigma}_s^2 = \mathbb{E}[(Z_s - \mu_s)^2]$.

Based on sublinear expectation, Peng et al. (2023) introduced a new risk measurement VaR model G-VaR for the time series $\{Z_s\}_{s=1}^{T}$, where G-VaR is defined by the sublinear expectation $\mathbb{E}[I(Z_s \leq x)]$ from a family of distribution $\{F_\theta\}_{\theta \in \Theta}$, where $I(\cdot)$ is the indicator function,

$$\mathrm{G-VaR}_\alpha(Z_s) := -\inf\{x \in \mathbb{R} : \mathbb{E}[I(Z_s \leq x)] > \alpha\}. \tag{7}$$

The advantage of the model G-VaR (7) is that we can find a distribution such that

$$F(s,x) = \mathbb{E}[I(Z_s \leq x)].$$

When $x \leq 0$, an explicit solution of $F(s, x)$ is

$$F(s,x) = \frac{2\bar{\sigma}_s}{\bar{\sigma}_s + \underline{\sigma}_s} \Phi\left(\frac{x - \mu_s}{\bar{\sigma}_s}\right), \tag{8}$$

where $\Phi(\cdot)$ is the cumulative distribution function of the standard normal distribution.

Based on distribution $F$, the explicit formula of G-VaR model (7) is given as follows,

$$\mathrm{G-VaR}_\alpha^{\beta_s}(Z_s) = -\mu_s - \bar{\sigma}_s \Phi^{-1}\left(\frac{\bar{\sigma}_s + \underline{\sigma}_s}{2\bar{\sigma}_s}\alpha\right), \tag{9}$$

where $\beta_s = (\mu_s, \underline{\sigma}_s, \bar{\sigma}_s)$. In formula (9), we use $(\underline{\sigma}_s, \bar{\sigma}_s)$ to present the volatility uncertainty of $Z_s$ at time $s$, and if $\underline{\sigma}_s = \bar{\sigma}_s$ it is the classical VaR model under classical normal distribution. The parameters are computed via the moving block method. As there are constraints on the window selection, we have devised an adaptive learning method to automatically modify the window length, namely ALM-G-VaR model.

For a given risk level $\alpha$, and parameter $\hat{\beta}_s = (\hat{\mu}_s, \hat{\underline{\sigma}}_s, \hat{\bar{\sigma}}_s)$, the counting function

$$\Pi(\alpha,t) = \frac{\sum_{s=1}^{t} I\left(Z_s < \mathrm{G-VaR}_\alpha^{\hat{\beta}_s}(Z_s)\right)}{t}, \tag{10}$$

converges to $\alpha$ in probability 1 as $t \to \infty$. In practice, we need that $\Pi(\alpha,t)$ is near the risk level $\alpha$. Therefore, we consider to adjust the estimations $(\hat{\underline{\sigma}}_s, \hat{\bar{\sigma}}_s)$ based on the value of $\Pi(\alpha, s-1)$. Note that, if $\Pi(\alpha, s-1) < \alpha$, to improve the accuracy of $\Pi(\alpha, s)$, we need to add the value of $\hat{\underline{\sigma}}_s$ and reduce the value of $\hat{\bar{\sigma}}_s$. If $\Pi(\alpha, s-1) > \alpha$, we need to reduce the value of $\hat{\underline{\sigma}}_s$ and add the value of $\hat{\bar{\sigma}}_s$. That is,

$$\tilde{\underline{\sigma}}_s = \begin{cases} \hat{\underline{\sigma}}_s + \underline{\delta}_s, & \Pi(\alpha, s-1) < \alpha, \\ \hat{\underline{\sigma}}_s, & \Pi(\alpha, s-1) = \alpha, \\ \hat{\underline{\sigma}}_s - \underline{\delta}_s, & \Pi(\alpha, s-1) > \alpha, \end{cases} \tag{11}$$

and

$$\tilde{\bar{\sigma}}_s = \begin{cases} \hat{\bar{\sigma}}_s - \bar{\delta}_s, & \Pi(\alpha, s-1) < \alpha, \\ \hat{\bar{\sigma}}_s, & \Pi(\alpha, s-1) = \alpha, \\ \hat{\bar{\sigma}}_s + \bar{\delta}_s, & \Pi(\alpha, s-1) > \alpha. \end{cases} \tag{12}$$

In practical analysis, when $\Pi(\alpha, s-1) < \alpha$, we consider to reduce $n_0$ and add $n_1$, when $\Pi(\alpha, s-1) > \alpha$, we consider to add $n_0$ and reduce $n_1$,

$$\tilde{n}_0 = \begin{cases} n_0 - w_0, & \Pi(\alpha, s-1) < \alpha, \\ n_0, & \Pi(\alpha, s-1) = \alpha, \\ n_0 + w_0, & \Pi(\alpha, s-1) > \alpha, \end{cases}$$

and

$$\tilde{n}_1 = \begin{cases} n_1 + w_1, & \Pi(\alpha, s-1) < \alpha, \\ n_1, & \Pi(\alpha, s-1) = \alpha, \\ n_1 - w_1, & \Pi(\alpha, s-1) > \alpha. \end{cases}$$

Here, we use $w_0$ and $w_1$ to adjust $n_0$ and $n_1$ of the previous step, which should satisfy $1 \leq \tilde{n}_1 \leq \tilde{n}_0$. In general, we set $w_0 = w_1 = 1$.

Now, we recalculate $\Pi(\alpha, s)$ with new parameters $(\tilde{\underline{\sigma}}_s, \tilde{\bar{\sigma}}_s)$, and replace $\Pi(\alpha, s)$ with

$$\tilde{\Pi}(\alpha, s) = \frac{\sum_{r=1}^{s} I\left(Z_r < \mathrm{G\text{-}VaR}_\alpha^{\tilde{\beta}_r}(Z_r)\right)}{s}, \tag{13}$$

where $\tilde{\beta}_s = (\tilde{\mu}_s, \tilde{\underline{\sigma}}_s, \tilde{\bar{\sigma}}_s)$.

**Proposition 4.1.** Let the log return of stock pricing $Z_s$ satisfy

$$Z_s = \mu + Y_j, \quad (j-1)m < s \leq jm, \quad 1 \leq j \leq n, \tag{14}$$

where $\mu$ is a constant, and $Y_j$ satisfy a normal distribution $N(0, \sigma_j^2)$. The number of the element of each group $m$ is not given and the total number of data is $T = nm$.

Then, for a given time $s$, when $\{\tilde{n}_0, \tilde{n}_1\} = \{n_0 + w_0, n_1 - w_1\}$,

$\mathbb{E}\left[\tilde{\underline{\sigma}}_s^2\right] \leq \mathbb{E}\left[\hat{\underline{\sigma}}_s^2\right], \mathbb{E}\left[\tilde{\bar{\sigma}}_s^2\right] \leq \mathbb{E}\left[\hat{\bar{\sigma}}_s^2\right]$, when $\{\tilde{n}_0, \tilde{n}_1\} = \{n_0 - w_0, n_1 + w_1\}$,

$\mathbb{E}\left[\tilde{\underline{\sigma}}_s^2\right] \geq \mathbb{E}\left[\hat{\underline{\sigma}}_s^2\right], \mathbb{E}\left[\tilde{\bar{\sigma}}_s^2\right] \geq \mathbb{E}\left[\hat{\bar{\sigma}}_s^2\right]$.

**Proof:** For a given risk level $\alpha$ and initial lengths of moving block $n_0$ and $n_1$, we can obtain the estimators for the upper and lower volatility $(\hat{\underline{\sigma}}_s, \hat{\bar{\sigma}}_s)$ and subsequently calculate $\tilde{\Pi}(\alpha, s-1)$. If $\tilde{\Pi}(\alpha, s-1) \neq \alpha$, then we need to adjust $\hat{\underline{\sigma}}_s$ and $\hat{\bar{\sigma}}_s$ according to $n_0$ and $n_1$.

We first consider the case $\tilde{\Pi}(\alpha, s-1) > \alpha$, and need to add $n_0$ to $\tilde{n}_0$ and reduce $n_1$ to $\tilde{n}_1$, where $\{\tilde{n}_0, \tilde{n}_1\} = \{n_0 + w_0, n_1 - w_1\}$. The adjustment estimators of volatility are:

$$\tilde{\underline{\sigma}}_s^2 = \min_{1 \leq j \leq \tilde{n}_0 - \tilde{n}_1 + 1} \tilde{\sigma}_{s,j}^2 \quad and \quad \tilde{\bar{\sigma}}_s^2 = \max_{1 \leq j \leq \tilde{n}_0 - \tilde{n}_1 + 1} \tilde{\sigma}_{s,j}^2,$$

where

$$\tilde{\sigma}_{s,j}^2 = \frac{\sum_{i=1}^{\tilde{n}_1} \left(Z_{s-\tilde{n}_0+i+j-1} - \tilde{\mu}_{s,j}\right)^2}{\tilde{n}_1 - 1},$$

and

$$\tilde{\mu}_{s,j} = \frac{\sum_{i=1}^{\tilde{n}_1} Z_{s-\tilde{n}_0+i+j-1}}{\tilde{n}_1}.$$

Note that $\{\tilde{n}_0, \tilde{n}_1\} = \{n_0 + w_0, n_1 - w_1\}$, the estimators $(\tilde{\underline{\sigma}}_s, \tilde{\bar{\sigma}}_s)$ can be expressed as:

$$\tilde{\underline{\sigma}}_s^2 = \min_{1 \leq j \leq n_0 - n_1 + 1 + w_0 + w_1} \tilde{\sigma}_{s,j}^2 \quad and \quad \tilde{\bar{\sigma}}_s^2 = \max_{1 \leq j \leq n_0 - n_1 + 1 + w_0 + w_1} \tilde{\sigma}_{s,j}^2.$$

Based on $\{n_0, n_1\}$, the estimators $(\hat{\underline{\sigma}}_s, \hat{\bar{\sigma}}_s)$ can be expressed as:

$$\hat{\underline{\sigma}}_s^2 = \min_{1 \leq j \leq n_0 - n_1 + 1} \hat{\sigma}_{s,j}^2 \quad and \quad \hat{\bar{\sigma}}_s^2 = \max_{1 \leq j \leq n_0 - n_1 + 1} \hat{\sigma}_{s,j}^2.$$

Without loss of generality, we assume $\hat{\sigma}_{s,j_0}^2 = \hat{\underline{\sigma}}_s^2$, where $\hat{\sigma}_{s,j_0}^2$ is calculated by $\{Z_{s-n_0+i+j_0-1}\}_{i=1}^{n_1}$. Note that the sequence $\{\tilde{\sigma}_{s,j}^2\}_{j=j_0}^{j_0+w_1}$ is also calculated by $\{Z_{s-n_0+i+j_0-1}\}_{i=1}^{n_1}$, and $\tilde{n}_0 \leq 2m$, it follows that

$$\min_{j_0 \leq j \leq j_0+w_1} \mathbb{E}\left[\tilde{\sigma}_{s,j}^2\right] \leq \mathbb{E}\left[\hat{\sigma}_{s,j_0}^2\right],$$

which deduces that

$$\mathbb{E}\left[\tilde{\underline{\sigma}}_s^2\right] \leq \mathbb{E}\left[\hat{\underline{\sigma}}_s^2\right].$$

In a similar manner, we can show that

$$\mathbb{E}\left[\tilde{\bar{\sigma}}_s^2\right] \geq \mathbb{E}\left[\hat{\bar{\sigma}}_s^2\right].$$

This completes the proof. □

**Assumption 2.2.** Let the parameter $\left(\hat{\underline{\sigma}}_s, \hat{\bar{\sigma}}_s\right)$ satisfy

$$\underline{l} < \inf_{s \geq 1} \hat{\underline{\sigma}}_s, \quad \sup_{s \geq 1} \hat{\bar{\sigma}}_s < \overline{l}.$$

Combining with Assumption 2.1, we have that the parameters $\left(\hat{\underline{\sigma}}_s, \hat{\bar{\sigma}}_s\right)$ and the adjustment parameters $\left(\tilde{\underline{\sigma}}_s, \tilde{\bar{\sigma}}_s\right)$ take value in the interval $[\underline{l}, \overline{l}]$. To evaluate the predictive performance of the ALM-G-VaR model, we first establish the convergence result for $\tilde{\Pi}(\alpha, s)$.

**Theorem 4.1.** Let Assumptions 2.1 and 2.2 hold. We have that

$$\lim_{t \to \infty} |\mathbb{E}[\tilde{\Pi}(\alpha, t)] - \alpha| = 0.$$

**Proof:** For a given risk level $\alpha$, by the formula of $\tilde{\Pi}(\alpha, t)$, it follows that,

$$\mathbb{E}[\tilde{\Pi}(\alpha, t)] = \frac{1}{t} \sum_{s=1}^{t} \mathbb{E}\left[I\left(Z_s < \text{G}-\text{VaR}_\alpha^{\tilde{\beta}_s}(Z_s)\right)\right].$$

Note that, $F(t, x) = \mathbb{E}\left[I(Z_t \leq x)\right]$, and

$$F(t, x) = \frac{2\bar{\sigma}_t}{\bar{\sigma}_t + \underline{\sigma}_t} \Phi\left(\frac{x - \mu_t}{\bar{\sigma}_t}\right).$$

Then, we have

$$\mathbb{E}[\tilde{\Pi}(\alpha, t)] = \frac{1}{t} \sum_{s=1}^{t} F\left(s, \text{G}-\text{VaR}_\alpha^{\tilde{\beta}_s}(Z_s)\right),$$

where

$$\text{G-VaR}_\alpha^{\check{\beta}_s}(Z_s) = -\tilde{\mu}_s - \tilde{\bar{\sigma}}_s \Phi^{-1}\left(\frac{\tilde{\bar{\sigma}}_s + \tilde{\underline{\sigma}}_s}{2\tilde{\bar{\sigma}}_s}\alpha\right).$$

Note that, $\Phi^{-1}(\cdot)$ is uniformly continuous in an any given bounded interval. For a given sufficiently small $\varepsilon > 0$, from the definition of sequence $(\tilde{\underline{\sigma}}_s, \tilde{\bar{\sigma}}_s)_{1 \leq s}$ in (11) and (12), and Assumption 2.2, there exist sequence $(\underline{\delta}_s, \bar{\delta}_s)_{1 \leq s}$, and $\delta_1(\varepsilon), \delta_2(\varepsilon) > 0$ such that

$$\delta_1(\varepsilon) < \left|\text{G-VaR}_\alpha^{\tilde{\beta}_s}(Z_s) - \text{G-VaR}_\alpha^{\hat{\beta}_s}(Z_s)\right| < \delta_2(\varepsilon),$$

and thus

$$\frac{\varepsilon}{2} < \left|F\left(s, \text{G-VaR}_\alpha^{\tilde{\beta}_s}(Z_s)\right) - F\left(s, \text{G-VaR}_\alpha^{\hat{\beta}_s}(Z_s)\right)\right| < \varepsilon.$$

Let us first consider the time $t_1 > 1$. If $\mathbb{E}\left[\tilde{\Pi}(\alpha, t_1)\right] > \alpha$, we reduce the value of $\hat{\underline{\sigma}}_{t_1+1}$ and add the value of $\hat{\bar{\sigma}}_{t_1+1}$, and obtain adjustment parameter $(\tilde{\underline{\sigma}}_{t_1+1}, \tilde{\bar{\sigma}}_{t_1+1})$. By the inequality above, it follows that

$$\frac{\varepsilon}{2} < \left|F\left(t_1+1, \text{G-VaR}_\alpha^{\tilde{\beta}_{t_1+1}}(Z_{t_1+1})\right) - F\left(t_1+1, \text{G-VaR}_\alpha^{\hat{\beta}_{t_1+1}}(Z_{t_1+1})\right)\right| < \varepsilon.$$

Then, there exists $t_2 > t_1$ such that $F\left(t_2, \text{G-VaR}_\alpha^{\tilde{\beta}_{t_2}}(Z_{t_2})\right) < \alpha$ and

$$\alpha - \frac{1}{t_2} \leq \mathbb{E}\left[\tilde{\Pi}(\alpha, t_2)\right] \leq \alpha.$$

Similarly, we can find $t_3 > t_2$ such that $F\left(t_3, \text{G-VaR}_\alpha^{\tilde{\beta}_{t_3}}(Z_{t_3})\right) > \alpha$, and

$$\alpha \leq \mathbb{E}\left[\tilde{\Pi}(\alpha, t_2)\right] \leq \alpha + \frac{1}{t_3}.$$

Therefore, there exists $N_0 > 0$ such that when $t > N_0$, and thus

$$|\mathbb{E}[\tilde{\Pi}(\alpha, t)] - \alpha| \leq \frac{1}{t}.$$

Using a similar manner with $\mathbb{E}\left[\tilde{\Pi}(\alpha, t_1)\right] < \alpha$, there exists $N_1 > 0$ such that when $t > N_1$,

$$|\mathbb{E}[\tilde{\Pi}(\alpha, t)] - \alpha| \leq \frac{1}{t}.$$

This completes the proof. $\square$

In the following, we show the convergence results of the violation rate in empirical analysis. Figure 6 shows the violation of S&P500 Index during January 1, 2007 to December 31, 2009 under the risk level $\alpha = 0.05$, which should converge to $\alpha$ in probability 1 according to Theorem 4.1. Note that, after short-term adjustment, the violation rate is stable, which shows that the ALM-G-VaR model has high accuracy in risk measurement.

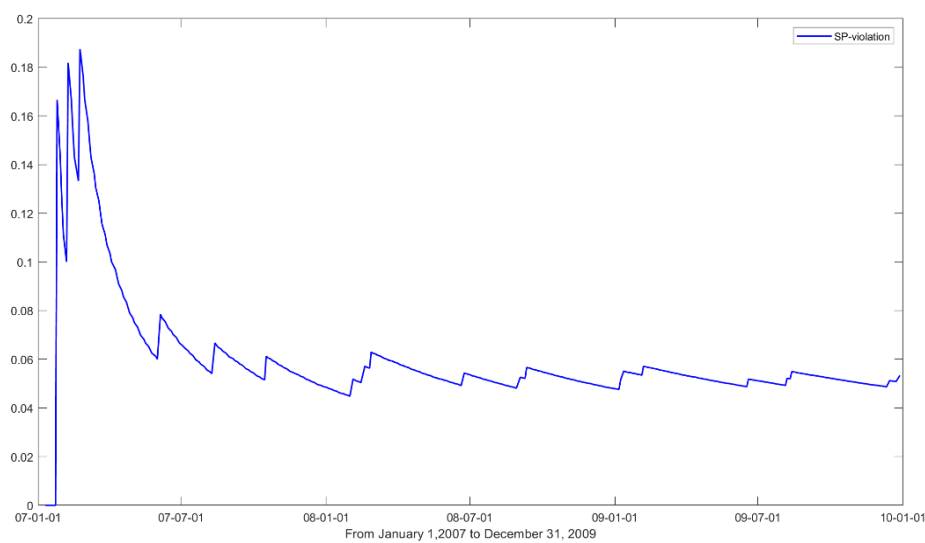

Figure 6 The violations of S&P500 Index under the risk level $\alpha = 0.05$

We use the test of a likelihood ratio for a Bernoulli trial and the test of a Christofferson independent to evaluate the predictive performance of the ALM-G-VaR model. We conclude the testing results of stock indexes in Table 5 with $\alpha = 0.05$. In Table 5, we calculate five indicators: Theo-Viol, Fact-Viol, $\hat{\alpha}$, $LR_{uc}$ and $LR_{ind}$, where Theo-Viol and Fact-Viol respectively represent the value of violation in theory and $\hat{\alpha}$ is the sample violations rate, $LR_{uc}$ denotes the likelihood ratio test statistics, $LR_{ind}$ denotes the Christofferson independent test statistics. In the performance of the indicators, Fact-Viol is close to Theo-Viol and $\hat{\alpha}$ is close to the risk level $\alpha = 0.05$, which indicates that ALM-G-VaR model can excellently capture the risk of the log return of the stock index. As for the test results, if the test statistic larger than the risk level α, that is $LR_{uc} > 0.05$, $LR_{ind} > 0.05$, then it implies the likelihood ratio test statistics and the Christofferson independent test statistics are passed under the confidential level 95%. According to the values of $LR_{uc}$ and $LR_{ind}$, we can conclude that almost all of stock indexes pass the two tests under the confidential level 95%. Therefore, the new ALM-G-VaR model has perfect predictive performance.

Table 5 Testing of S&P 500 Index with $\alpha = 0.05$

| Index | Theo-Viol | Fact-Viol | $\hat{\alpha}$ | $LR_{uc}$ | $LR_{ind}$ |
|---|---|---|---|---|---|
| S&P500 | 18.75 | 19 | 0.0507 | 0.9529 | 0.3250 |
| IXIC | 18.75 | 19 | 0.0507 | 0.9529 | 0.0726 |
| FTSE | 18.75 | 18 | 0.0480 | 0.8581 | 0.0527 |
| CSI300 | 18.75 | 21 | 0.0560 | 0.6006 | 0.4246 |
| GDAXI | 18.75 | 18 | 0.0480 | 0.8581 | 0.8826 |